\documentclass[amsmath,amssymb,superscriptaddres,preprints]{revtex4-2}
\usepackage[latin1]{inputenc}
\usepackage{amsmath}
\usepackage{amsfonts}
\usepackage{amssymb}
\usepackage{makeidx}
\usepackage{graphicx,epsfig}
\usepackage{epstopdf}
\usepackage{dcolumn}
\usepackage{comment}
\usepackage{caption}
\usepackage{subcaption}
\usepackage{epstopdf}
\usepackage{dcolumn}
\usepackage[colorlinks=true, citecolor=blue, urlcolor = blue, linkcolor= red, bookmarks=true]{hyperref}

\begin{document}

\title{Photon orbits and phase transition for Letelier AdS black holes immersed in perfect fluid dark matter}
\author{Ashima Sood}
 \email{ashimasood1993@gmail.com}
 \affiliation{Department of Mathematics, Netaji Subhas University of Technology, New Delhi-110078, India}
\affiliation{Centre for Theoretical Physics, Jamia Millia Islamia, New Delhi 110025, India}
\author{Md Sabir Ali}
\email{alimd.sabir3@gmail.com }
\affiliation{Department of Physics, Mahishadal Raj College, Purba Medinipur, West Bengal, 721628, India}
\affiliation{Institute of Theoretical Physics $\&$ Research Center of Gravitation, Lanzhou University, Lanzhou 730000, China}
\affiliation{Key Laboratory of Quantum Theory and Applications of MoE, Lanzhou University, Lanzhou 730000, China}
\affiliation{Lanzhou Center for Theoretical Physics $\&$ Key Laboratory of Theoretical Physics of Gansu Province, Lanzhou University, Lanzhou 730000, China} 
\author{J. K. Singh}
\email{jksingh@nsut.ac.in}
\affiliation{Department of Mathematics, Netaji Subhas University of Technology, New Delhi-110078, India}
\author{Sushant G. Ghosh} 
\email{sghosh2@jmi.ac.in}
\affiliation{Centre for Theoretical Physics, Jamia Millia Islamia, New Delhi 110025, India}
\affiliation{Astrophysics and Cosmology Research Unit, School of Mathematics, Statistics and Computer Science, University of KwaZulu-Natal, Private Bag X54001, Durban 4000, South Africa}

\begin{abstract}
  We obtain an exact solution of spherically symmetric Letelier AdS black holes immersed in perfect fluid dark matter (PFDM). Considering the cosmological constant as the positive pressure of the system and volume as its conjugate variable, we analyse the thermodynamics of our black holes in the extended phase space. Owing to the background clouds of strings parameter ($a$) and the parameter endowed with PFDM ($\beta$), we analyse the Hawking temperature, entropy and specific heat. We also investigate the relationship between the photon sphere radius and the phase transition for the Letelier AdS black holes immersed in PFDM. Through the analysis, we find with a particular condition, there are non-monotonic behaviours between the photon sphere radius, the impact parameter, the PFDM parameter, temperature, and pressure. We can regard both the changes of photon sphere radius and impact parameter before and after phase transition as the order parameter; their critical exponents near the critical point are equal to the same value 1/2, just like ordinary thermal systems. These indicate that a universal relation of gravity may exist near the critical point for a black hole thermodynamic system.  
\end{abstract}

\maketitle

\section{Introduction}
Black holes are one of the strangest objects in the universe that primarily reside at the heart of the galaxies. A well-defined boundary called the event horizon specifies them, separates the black hole's interior and exterior regions and is causally disconnected \cite{Wald:1999vt, Israel:1967wq}. The event horizon is also a null hypersurface. The thermodynamics of any generic black holes are studied in terms of the event horizon radius \cite{Carlip:2014pma}. The thermodynamics of black holes connect several interdisciplinary branches of theoretical physics, e.g., general relativity, quantum mechanics, statistical mechanics, and information theory. The first study that revolutionized black hole thermodynamics was Bardeen \textit{et. al.} \cite{Bardeen:1973gs}. They formulated the laws of black hole mechanics in analogy with the ordinary thermodynamic system in their pioneering work. The continued interests led Bekenstein to propose that the entropy of the black holes must have a direct correspondence to the horizon area itself \cite{Bekenstein:1972tm, Bekenstein:1973ur}. The further proposal of Hawking and Page \cite{Hawking:1982dh} that the black holes in anti-de Sitter (AdS) spacetime must exhibit a phase transition between the pure thermal AdS bath and the Schwarzschild-AdS spacetimes, familiarly known as the Hawking-Page phase transitions. It triggered the investigations of the black holes thermodynamics in AdS spacetimes at the forefront of modern scientific research because of its relevance to AdS/CFT correspondence \cite{Kastor:2009wy, Dolan:2011xt, Dolan:2010}. The AdS/CFT correspondence tremendously affects the thermodynamics and phase structure of various AdS black holes. The Reissner-N{{o}}rdstr$\Ddot{o}$m-AdS black holes exhibit a van der Waals-like phase transition \cite{Chamblin:1999tk, Chamblin:1999hg, Wang:2019cax}. When we model the charged-AdS black holes as a canonical ensemble, we find a first-order phase transition that terminates at the second order where critical parameters are determined through the $P-v$ criticality \cite{Wang:2019cax}. Such black holes also show a Hawking-Page-like phase transition when considered a grand canonical ensemble. A zeroth-order phase transition, the reentrant one, also seems to exist for specific AdS black hole systems, e.g., the Born-Infeld-AdS black holes \cite{Dehyadegari:2017hvd,
NaveenaKumara:2020biu}. In all such analyses, one considers the cosmological constant as a thermodynamic pressure $ P=(-\Lambda/8\pi) $ and its conjugate quantity as a thermodynamic volume. We also observed multiple exotic phenomena, including the coexistence lines in the $P-T$ diagrams, terminating at the finite critical points and the $P-v$ criticality mimicking the van der Waals type phase transition \cite{Li:2022vcd}. Despite such vivid descriptions, the thermodynamics of AdS black holes remains puzzling. We expose the phase structure through various perspectives. For example, some researchers probed the microstructure of the black holes through the investigations of the related Ruppeiner geometry \cite{Li:2022vcd, NaveenaKumara:2020lgq,
NaveenaKumara:2020biu, Hegde:2021hjq}. We can construct the repulsive or attractive nature of the micro-molecules, through such formalism as reported in \cite{Wei:2020poh, Wei:2020cqn, Wei:2019ctz, Wei:2019uqg, Zhou:2020vzf, Dehyadegari:2020ebz}. The thermodynamic phase structure of the black holes in AdS spacetimes is exposed to probe some of the observational signatures, such as the geodesics structure of the circular photon orbits \cite{Wei:2017mwc, NaveenaKumara:2019nnt}, and the black hole shadow radius \cite{Luo:2023ndw, Guo:2022yjc, Zou:2017juz}, and the quasinormal modes \cite{Lan:2020fmn, Abbas:2023pug, Singh:2022xgi, Li:2017kkj, Liu:2014gvf}. The discontinuous changes of the phase transition behaviour between the small black hole (SBH) and the large black hole (LBH) may help determine the order parameter of various thermodynamic quantities of physical interests.\\

Motivated by the correlation of the dynamics and the thermodynamics, in the context of the AdS black holes geometry, a well-posed attempt has recently been put into the analysis relating to gravity and thermodynamics. The phase transition behaviour between SBH-LBH stable phases is probed through the unstable photon orbit radius. The phase transition behaviour of the unstable photon orbit radius and the minimum impact parameter have oscillatory behaviour similar to the $T-r_h$ and the $T-S$ diagrams of the usual vdW-like fluids. Another motivation comes from the discontinuous changes of the photon orbit radius and the minimum impact parameter between SBH-LBH phases, with a critical exponent of $1/2$. Therefore, the phase transition behaviour is scrutinized for many AdS black hole spacetimes. The connection of the unstable photon circular orbits in several other contexts has also been significantly studied. Keeping the utility of the phase transition behaviour, we probe them through the photon orbit radius and the minimum impact parameter. In the present paper, we propose a possible correlation connecting the unstable circular null geodesics and the minimum impact parameter for the Letelier black holes in the AdS spacetime in the presence of perfect fluid dark matter (PFDM). The PFDM spacetime in AdS background has been studied in different contexts in order to investigate, e.g., the topological classes of thermodynamic phase transitions, the justification of the weak censorship conjecture and many more \cite{Rizwan:2023ivp, Li:2022fmq, Xu:2017bpz}. The criticality and thermal phase transition behaviour is explored in string cloud background in the presence of additional field \cite{Rahmani:2024cfi}. Another study is also devoted to the shadow analysis for the Letelier spacetime \cite{Sun:2024xtf, Vishvakarma:2023tnl} while the others are studied in another contexts \cite{Liu:2023kxd, Guo:2023zwy, Yang:2022ryf, Rodrigues:2022zph, Mustafa:2022xod, He:2021aeo, Liang:2020uul, Toledo:2020xxc, Li:2023ztd, Zhai:2023wig, Yang:2023agi, Alipour:2023css,Ama-Tul-Mughani:2023ehc}. The Letelier spacetime in a spherically symmetric case is obtained when a gauge invariant model of an array of a cloud of strings coupled to gravity in the framework of Einstein's general relativity. The enigmatic phenomenon of Letelier AdS, black holes in PFDM, may find its origin in the presence of a surrounding fluid-like dark matter, a pivotal and unanswered inquiry within the realm of physics. Dark matter is 25\% of the universe's energy density. Our investigation holds the potential and might offer illuminating glimpses into the elusive properties of dark matter. \\

After obtaining the exact solution, we investigate the Hawking temperature, the $P-v$ criticality, the Gibbs free energy and other thermodynamic quantities of physical interest for the Letelier black holes in AdS spacetime surrounded by PFDM. We also explicitly show the effects of the cloud of strings parameter and the parameter characterizing the PFDM on the photon orbits and related quantities by plotting them numerically. \\

We organized the paper as follows. Section \textcolor{blue}{I} we briefly review the Letelier black holes in the AdS spacetimes surrounded by PFDM. Section \textcolor{blue}{II} discusses the black holes' usual phase transition phenomena, compares them with the typical van der Waals-like fluid, and distinguishes the SBH and LBH phases. In the next section, we investigate the unstable photon orbit radius and the minimum impact parameter using the effective potential. We also get the discontinuous changes of the photon orbits and the impact parameter at the point where SBH-LBH phase transitions occur, acting as an order parameter in \textcolor{blue}{IV}. Section \textcolor{blue}{V} deals with the null geodesics and the phase transition behaviour near the critical points. Finally, we conclude the paper in Section \textcolor{blue}{VI}.

\section{Black hole solution immersed in PFDM surrounded by clouds of strings
} \label{solution}
The action for ($3+1$) dimensional gravity with cosmological constant
\begin{equation}\label{action}
    \frac{1}{16\pi} \int d^{4}x \sqrt{-g}\left[R +6l^{-2}+\mathcal{L_{DM}}\right]+I_{S},
\end{equation}
where $R$ is the Ricci scalar, $g$ is the determinant of the metric tensor $g_{\mu\nu}$, $\mathcal{L_{DM}}$ gives the Lagrangian density of PFDM and \cite{Letelier:1979ej} $$\mathcal{I}_{S} = \int_{\Sigma} \mathcal{L} d\lambda^{0}d\lambda^{1}$$ is the Nambu-Gotu action of a string characterised by timelike and spacelike parameters ($\lambda^{0},\lambda^{1}$) respectively. The Lagrangian density of string cloud is $\mathcal{L}=m(\gamma)^{-1/2}$ with $m$ a dimensionless positive constant representing each string. The worldsheet of the string has an induced metric \cite{Ghosh:2014dqa,Singh:2020nwo} $$\gamma_{ab}=g_{\mu\nu}\frac{dx^{\mu}}{dx^{a}}\frac{dx^{\nu}}{dx^{b}}$$ with $|\gamma|=$ det $\gamma_{ab}$. The bivector associated with the string worldsheet has the form $$\sigma^{ab}=\epsilon^{ab}\frac{dx^{\mu}}{d\lambda^{a}}\frac{dx^{\nu}}{d\lambda^{b}},$$ where $\epsilon^{0~1}=-\epsilon^{1~0}=1$ is the $2$ dimensional Levi-Civita tensor. Hence the Lagrangian density becomes $L=m\left[-1/2\sigma^{\mu\nu}\sigma_{\mu\nu}\right]$. Since $T^{\mu\nu}=\partial\mathcal{L}/\partial g^{\mu\nu}$, the energy momentum tensor for clouds of strings is $T^{\mu\nu}=\rho\sigma^{\mu\Sigma}\sigma^{\nu}_{\Sigma}/(-\gamma^{1/2})$ where $\rho$ is the proper density of the string cloud. 

Extremizing the action leads to the Einstein field equation \cite{Ghosh:2020ijh}
\begin{equation}\label{ee}
    G_{ab}-\frac{3}{l^{2}}g_{ab}=8\pi(T^{S}_{ab}+T^{DM}_{ab}),
\end{equation}
where $T^{S}_{ab}$ and $T^{DM}_{ab}$, respectively, are energy-momentum tensors of the cloud of strings and perfect fluid dark matter. In the rest frame associated with the observer $\rho=a/r^2$ and the components of energy-momentum tensor for clouds of strings \cite{Sood:2022fio,Kumar:2023gjt}
\begin{equation}\label{tl}
    (T^{a}_{b})^{S}=\mbox{diag}\left(-\frac{a}{r^2},\frac{a}{r^2},0,0\right),
\end{equation}
where $a$ is some real constant. The energy-momentum of PFDM \cite{Zhang:2020mxi} reads
\begin{equation}\label{tpfdm}
    (T^{a}_{b})^{DM}=\mbox{diag}\left(-\epsilon,P_{r},P_{\theta},P_{\phi}\right),
\end{equation}
with 
\begin{equation}
    \epsilon = -P_{r} = -\frac{\beta}{8\pi r^3},\quad\quad P_{\theta} = P_{\phi} = -\frac{\beta}{16\pi r^3},
\end{equation}
where $\epsilon$ is the energy density and $P_{r}$ is the radial and $P_{\theta}$, $P_{\phi}$ are the tangential pressures of the PFDM. Using Eq. (\ref{ee}) with (\ref{tl}) and (\ref{tpfdm}), and the static spherically symmetric metric \cite{Kumar:2023gjt}
\begin{equation}\label{met}
    ds^{2}=-f(r)dt^2+\frac{1}{f(r)}dr^2+r^2 \left(d\theta^2\,+\,sin^2\,\theta\, d\phi^2\right),
\end{equation}
For this system, the Letelier black holes immersed in PFDM with a negative cosmological constant for the metric (\ref{met}) reads :
\begin{equation}\label{sol}
    f(r)=1-\frac{2m}{r}-a+\frac{\beta}{r}\ln\left(\frac{r}{|\beta|}\right)+\frac{r^2}{l^2}.
\end{equation}
Here $m$ arising from integration is related to black hole mass, and $\beta$ is the PFDM parameter. Henceforth, we shall refer to our solution (\ref{sol}) as Letelier AdS black holes in PFDM. Our solution extends Letelier black holes in AdS spacetime ($\beta=0$) to include PFDM. When $ \beta=a=0$, we obtain Schwarzchild AdS black holes.  
\paragraph{Energy Conditions:} We examine the status of the energy conditions \cite{Hawking:1973uf, Kothawala:2004fy, Ghosh:2008zza} and assess their applicability to our solution (\ref{met}). The components of the stress-energy tensor $T_{ab}$, calculated using Eq. (\ref{ee}), are then 
\begin{equation}\label{rho}
\begin{aligned}
    \rho=\frac{ar-\beta }{r^{3}}  = -P_{r},\\
    P_{\theta}=P_{\phi}=-\frac{\beta}{2r^3}.
\end{aligned}
\end{equation}
\begin{itemize}
    \item The weak energy condition (WEC) requires that for any timelike vector $t^{a}$, $T_{ab}t^{a}t^{b}\geq 0$ everywhere, that is $\rho \geq 0$ and $\rho + P_{\theta}\geq 0$
It turns out that 
    \begin{equation}
    \rho+P_{\theta}=\frac{2ar-3\beta}{2r^2},    
    \end{equation}
    implies that $\rho+P_{\theta}$ is always positive for $r>0$ and $0<a<1$.
    The WEC is satisfied when
$  {(ar-\beta)}/{r^3}\geq 0 $ for  $0<a<1$.  
  \item The null energy condition (NEC) requires that $T_{ab}t^{a}t^{b}\geq 0$ holds throughout the entire space-time for any null vector $t^{a}$. Consequently, note that  $\rho+P_{r}\geq0$ is inherently zero. Additionally, $\rho+P_{\theta}$ ($P_{\phi}$) $\geq0$ is consistently satisfied for $0<a<1$ and $r>0$.
    \item The strong energy condition (SEC) states that $T_{ab}t^{a}t^{b}\geq 1/2T^{a}_{a}t^{b}t_{b}$ globally, for any timelike vector $t^{a}$, that is 
    \begin{equation}\label{secc}
        \begin{aligned}
        \rho+P{r}+2P_{\theta}=\frac{(2ar-5\beta)}{2r^3}.
        \end{aligned}
    \end{equation}
    \end{itemize}
    Thus, it can be seen from Eq. (\ref{secc}) that the SEC is satisfied.

\begin{figure}[h!h!] 
\begin{tabular}{c c }
\includegraphics[width=0.5\textwidth]{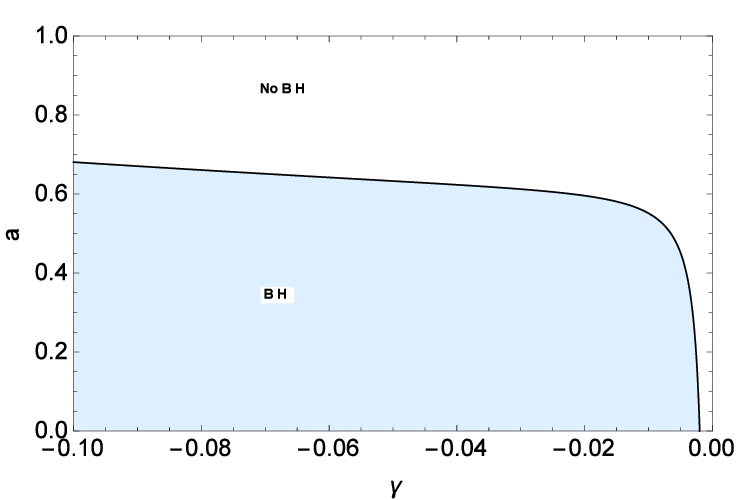}
\end{tabular}
\caption{ Parameter space for the Letelier black holes in PFDM showing the possible values of $a$ and $\gamma=\beta/l$.}
\label{fig:para}
\end{figure}

\begin{widetext}
\begin{figure}[h!h!] 
\begin{tabular}{c c }
\includegraphics[width=0.5\textwidth]{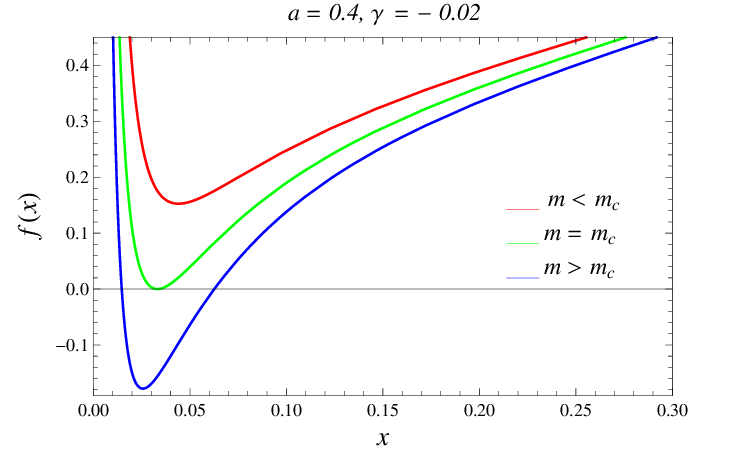}
\includegraphics[width=0.5\textwidth]{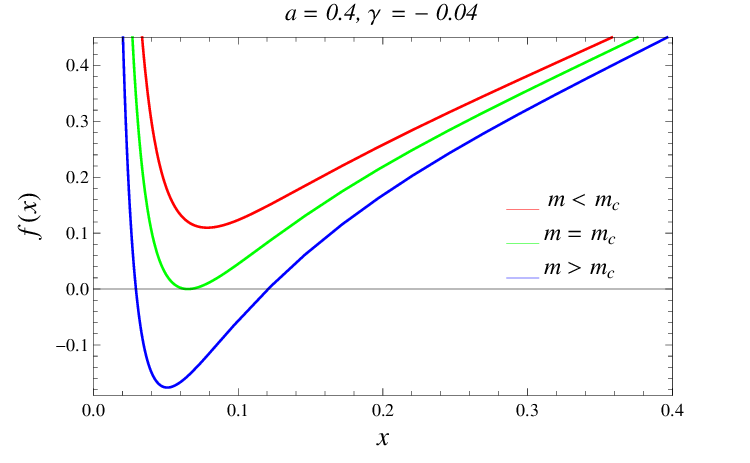}
\end{tabular}
\caption{ The plot of metric function $f(x)$ vs $x=r/l$. In the degenerate cases, we get an extremal black hole for $ m=M_c=0.004910$ (left) and $M=M_c=0.009926 $ (right). A naked singularity is evident for $ M < M_c $.}
\label{horizon1}
\end{figure}
\end{widetext}

\paragraph{Extremal Black Holes:} The horizons are determined by the points where the metric's singularity occurs, characterized by $f(r) = 0,"$ indicating coordinate singularities. As illustrated in Fig. \ref{horizon1}, solving the equation $f(x) = 0$ numerically reveals the possible existence of two roots. Corresponding to specific values of $a$ and $\gamma=\beta/l$, we identify the mass parameter's critical point ($M_{c}$). Consequently, for mass values ($M>M_{c}$), we encounter two zeros representing the Cauchy and event horizons (refer to Fig. \ref{horizon1}). The plot in Fig. \ref{horizon1} illustrates the Cauchy and event horizons for $a=0.4$ and $\gamma=-0.02$($-0.04$). Additionally, the plot indicates the absence of a horizon (naked singularity) for $M<M_{c}$. However, when $M=M_{c}$, an extremal black hole with a degenerate horizon emerges. The expression for extremal black holes is derived from the equation
\begin{equation}
    f(r)=0=\frac{\partial f(r)}{\partial r},
\end{equation}
which yields the values of $r_{o}$ and $M_{c}$ as
\begin{equation}
    r_{o}=\frac{1}{3}\left[\frac{-2(1-a) l^2 +(2~\mathcal{U}^{2})^{\frac{1}{3}}}{(2^2~\mathcal{U})^{\frac{1}{3}}}\right]
\end{equation}
and 
\begin{widetext}
\begin{equation}
    M_{c}=\frac{1}{18~\mathcal{U}^{\frac{1}{3}}}\left[-2^{\frac{4}{3}}(a-1)^2-3\beta~\mathcal{U}^{\frac{1}{3}} + {2~\mathcal{U}}^{\frac{2}{3}}-{2~\mathcal{U}}^\frac{2}{3}a+9\beta~\mathcal{U}^\frac{1}{3}\ln{(\frac{2^\frac{1}{3}~\mathcal{U}^\frac{2}{3}-2(a-1)l^2}{2^\frac{2}{3}~\mathcal{U}^\frac{1}{3}~3|\beta|})}\right]
\end{equation}
\end{widetext}
where
\begin{equation}
    \mathcal{U}=-9~l^2~\beta+\sqrt{-4(-1+a)^3 l^6 +81 l^4 \beta^2}
\end{equation}
In Fig.~\ref{fig:para}, we present the parametric space for the values of $a$ and $\gamma$ while keeping the mass constant. The solid black curve in the figure signifies the extremal black hole, and the blue region denotes the presence of black holes with two horizons.  Conversely, the white region represents the absence of a black hole. The thermodynamic analysis of the Letelier spacetime immersed in PFDM pertains to values of the CS and PFDM parameters, $a$ and $\gamma$, falling within the region associated with a black hole.

\section{Thermodynamics and $P-V$ criticality }\label{thermo}
\begin{figure}[h!] 
\includegraphics[width=0.5\textwidth]{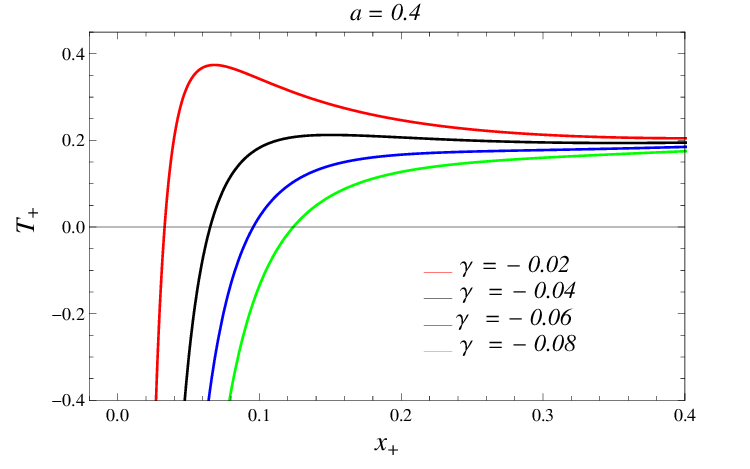}
\caption{The Hawking temperature $T_{+}$ vs horizon $x_+ = r_+/l$ for $a=0.4$ and different values of $\gamma=\beta/l$.} 
\label{fig:temp}
\end{figure}

Next, we try to connect Letelier in PFDM black holes with the concept of black hole chemistry \cite{Zhang:2020mxi}, which requires the study of the thermodynamic properties of Letelier in PFDM in AdS background. To investigate chemical aspects, we must discuss all thermodynamic variables \cite{Kubiznak:2012wp}. The relation $f(r_+)=0$ gives the ADM mass of the black hole solution and reads  
\begin{equation}\label{mass}
M_{+}=\frac{1}{2}\left[(1-a)r_{+}+\beta \ln{\frac{r_{+}}{|\beta|}}+\frac{r_{+}^3}{l^2}\right].  
\end{equation}
The identification of mass as chemical enthalpy, which is also the system's total energy, originates from consideration of the cosmological constant ($\Lambda$) as the system's pressure. 
The temperature of the black hole is viewed as radiation emitted by a black body satisfying the relation \cite{Kubiznak:2016qmn} $T={\kappa}/{2\pi}$, where $\kappa$ is the surface gravity of the black hole. The expression for the temperature is obtained as 
\begin{equation}\label{temp}
T_{+} = \frac{1}{4\pi} {\left[\,\frac{3}{l^2}-\frac{(1-a)r_{+}+2\beta}{r_{+}^{3}}\,\right]}.
\end{equation}
The local extrema of Hawking temperature are identified from the expression $Z(r_{+})=\partial T_{+}/\partial r_{+}$. Numerically, $Z(r_{+})$ yields a pair of complex roots and one real root as the solution of $Z(r_{+})=0$, which corresponds to the extrema of the Hawking temperature. Fig.~\ref{fig:temp} shows the plot of temperature vs horizon radius varying the values of the dark parameter $\gamma =\beta/l$. For decreasing values of the ratio $\beta/l$ the extremal values of the temperature curve gradually merge to become monotonically increasing. Also, the temperature becomes zero (cf. Fig.~\ref{fig:temp}) at horizon radius $r_{+}=0.09562$ for $a=0.4$ and $\gamma=-0.06$. Further, the plot also indicates a positive slope for large and small black holes and a negative slope for intermediate black holes. This behaviour of the Hawking temperature is indicative of van der Waals-like first-order phase transition \cite{Kubiznak:2012wp, Belhaj:2013cva, He:2016fiz, Kastor:2009wy, Cvetic:2001bk,
Herscovich:2010vr, Lee:2015xlp}.
The entropy of the black hole is calculated from the Bekenstein area law \cite{Zhang:2020mxi}
\begin{equation}
    S_{+}=\frac{A}{4}=\int^{2\pi}_{0}\int^{2\pi}_{0}\sqrt{g_{\theta\theta}\,g_{\phi\phi}}\,d\theta \,d\phi=\pi r_{+}^2,
\end{equation}  
for Letelier AdS black holes in PFDM. The mass may be a function of three thermodynamic variables, e.g., the entropy $S$, pressure $P$, and PFDM parameter $\beta$. With the consideration of pressure in terms of the cosmological constant $\Lambda$, the first law of black hole thermodynamics is derived as the total derivative of the mass $M_{+}$ \cite{Kubiznak:2012wp,
Kubiznak:2016qmn}
\begin{equation}
    dM_{+}=\left(\frac{\partial M_{+}}{\partial S_{+}}\right)_{\beta, P}\,dS{_+}\,+\,\left(\frac{\partial M_{+}}{\partial P}\right)_{\beta, S_{+}}\,dP\, +\,\left(\frac{\partial M_{+}}{\partial \beta}\right)_{S_{+}, P}\,d\beta
\end{equation}
This identification of the black hole first law with that of the classical thermodynamics is possible after mass is viewed as enthalpy and a pressure term is accounted with the relation \cite{dr2} $P=-\Lambda/8\pi$. The conjugate quantities are defined as
\begin{equation}
    T_{+}=\left(\frac{\partial M_{+}}{\partial S_{+}}\right),~~~ V_{+}=\left(\frac{\partial M_{+}}{\partial P}\right),~~~ \pi=\left(\frac{\partial M_{+}}{\partial \beta}\right),
\end{equation}
giving the expression for the first law as \cite{Hansen:2016ayo}
\begin{equation}
    dM_{+}\,=\,T_{+}dS_{+}-P\,dV_{+}+\pi\,d\beta. 
    \label{firstlaw}
\end{equation}
The Smarr formula can be realized from the first law with the help of a small scaling argument and Euler's theorem. In the asymptotically flat case, the first law and Smarr relation are found out by geometrical means \cite{Kastor:2011qp,Dolan:2011xt}. Scaling of the cosmological constant by (length)$^{-2}$, mass $M_{+}$ by (length)$^1$ and area $A$ by (length)$^{2}$ in conjunction with Euler's theorem give the following expression for the Smarr relation \cite{Kastor:2009wy}     
\begin{equation}\label{smarr relation}
M_{+}=2\left(\frac{\partial M_{+}}{\partial A}\right)~A-2\left(\frac{\partial M_{+}}{\partial \Lambda}\right)~\Lambda
\end{equation}
where Area $A$ is perceived as entropy and $\Lambda$ as pressure in black hole thermodynamics. The presence of the PFDM parameter in our black hole configuration modifies the Smarr relation to \cite{Kastor:2009wy, Papadimitriou:2005ii}
\begin{equation}
    M_{+}=2T_{+}S_{+}-2PV_{+} +\Pi \beta
\end{equation}
The analytical expressions for the conjugate quantities $\Pi$ and volume of the black hole system is \cite{Zhang:2020mxi}
\begin{equation}
\Pi=\frac{1}{2}\left[-1+\ln\frac{r_{+}}{|\beta|}\right],\qquad V_{+}=\frac{4}{3} \pi r_{+}^3,\qquad P=\frac{3}{8 \pi l^2}.    
\end{equation}
In the limit $\beta \to 0$ the above quantities reduce to the corresponding expressions for the Letelier AdS black holes \cite{Sood:2022fio}. 

\begin{widetext}
\begin{figure}[h!h!] 
\begin{tabular}{c c }
\includegraphics[width=0.5\textwidth]{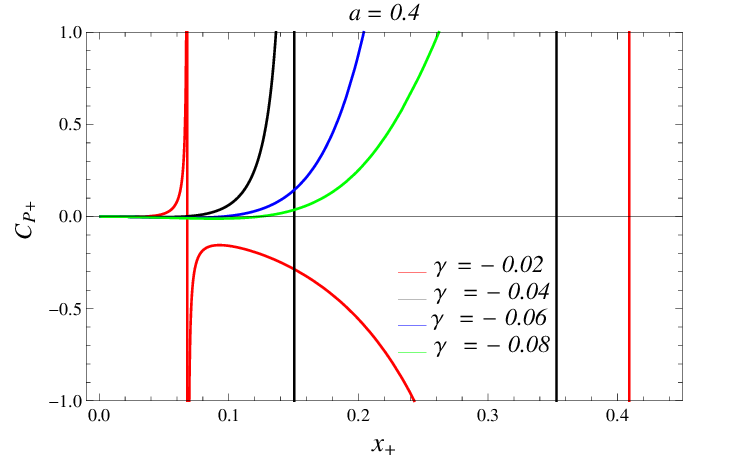}
\includegraphics[width=0.5\textwidth]{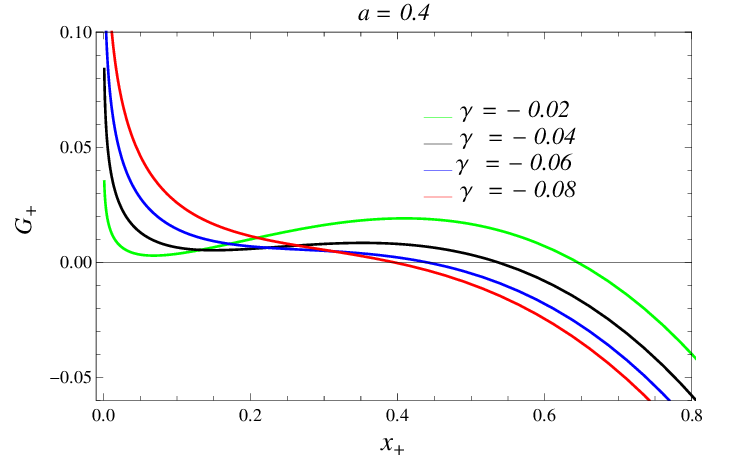}
\end{tabular}
\caption{The plot of specific heat (a) $C_{P+}$ vs horizon $x_+=r_{+}/l$ (left) (b) Gibbs free energy $G_{+}$ vs horizon $x_+=r_{+}/l$ (right) for decreasing values of $\gamma$.}
\label{fig : cp and gx} 
\end{figure}
\end{widetext}

\begin{widetext}

\begin{figure}
\begin{tabular}{c c }
\includegraphics[width=0.5\textwidth]{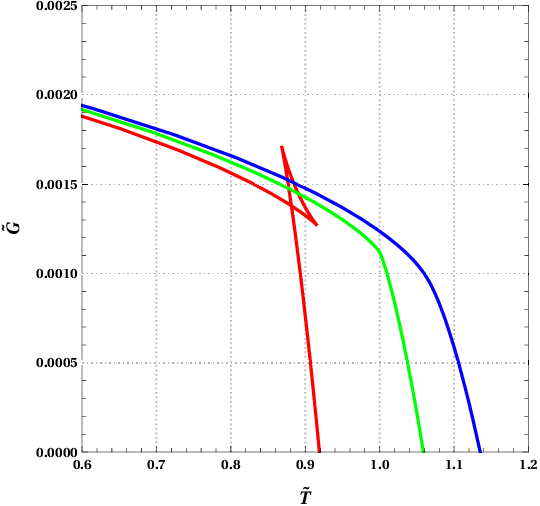}
\end{tabular}
\caption{The plot of Gibbs free energy $\Tilde{G}$ vs temperature $\Tilde{T}$ showing a swallowtail for $P<P_{c}$. The curve attains a cusp at $P=P_{c}$, which is also a point of second-order phase transition and further becomes monotonically increasing for $P>P_{c}$. }
\label{fig : gt}
\end{figure}

\end{widetext}

The local stability is determined by specific heat $C_V{_{+}}$ and $C_P{_{+}}$ computed from the expression $C_{V}{_{+}}=T{_{+}}(\partial S_{+}/\partial T_{+})_{V_+,\,\beta}$ and $C_{P}{_{+}}=T_{+}(\partial S_{+}/\partial T_{+})_{P,\,\beta}$ \cite{Singh:2022xgi,Ghosh:2014dqa}. It is well known that a positive value of the heat capacity signifies a locally stable black hole space-time. In contrast, a negative value means a locally unstable black hole solution. The specific heat at constant volume, $C_{V}{_{+}}$ comes out to be zero since entropy $S_{+}$ is interdependent on the volume $V_{+}$ \cite{Kubiznak:2012wp} for the solution under investigation. The specific heat at constant pressure, $C_{P}{_{+}}$ is   
\begin{equation}
    C_{P}{_{+}}=\frac{2\pi r_{+}^2\,(\,3r_{+}^3\,+\,(1-a)\,r_{+}\,+\beta)}{3r_{+}^3\,-l^2(r_{+}\,(1-a)\,-2\beta)}.
\end{equation}      
Fig.~\ref{fig : cp and gx} (left) shows the behaviour of $C_{P}{_{+}}$ for various values of PFDM parameter $\gamma=\beta/l$. The plot in Fig.~\ref{fig : cp and gx} shows that a second order phase transition is obtained at two critical points $r_{c1}$ and $r_{c2}$ for $\gamma=-0.02$($-0.04$). The critical points indicate the divergence of the specific heat as the values of $\gamma=\beta/l$ decrease. These points of inflexion are the points of extrema for the temperature. For $\gamma=-0.06$($-0.08$), the plot positively increases, signifying a stable black hole region.    
The global stability of the black hole system is measured by Gibbs free energy \cite{Ghosh:2020ijh}. The expression for free energy is computed from $G_{+}\,=\,M_{+}-T_{+}S_{+}$ which is 
\begin{equation}
\label{gibbs}
G_{+}=\frac{1}{4}\left[r_{+}\,(1-a)-\frac{r_{+}^2}{l^2}-\beta+2\,\beta\, \ln\,\frac{r_{+}}{|\beta|}\right]
\end{equation}
Fig.~\ref{fig : cp and gx} (right) and~\ref{fig : gt} determine the behaviour of the free energy $G$ in the $G-x$ plane as the ratio $\beta/l$ decreases. The right panel in Fig.~\ref{fig : cp and gx} shows the free energy becomes negative which signifies the presence of a Hawking-Page phase transition. The free energy curve becomes ``$0$" which transitions from pure radiation to a stable black hole configuration. 
Next, we show the plot of the Gibbs free energy vs temperature to highlight the small-to-large black hole transition. Fig.~\ref{fig : gt} shows a swallowtail-like structure for pressures below the critical pressure. The plot shows the existence of three states, i.e. the Small (SBH), large (LBH) and intermediate (IBH) black holes and the transitions between them. The red curve in the plot represents two first-order phase transitions between stable SBH to less stable IBH and IBH to stable LBH which can be seen where the Gibbs free energy intersects with itself. The figure indicates a swallowtail for values of pressure below the critical pressure $P_{c}$, and the behaviour of the free energy at the critical pressure is signified by the green curve, which shows a cusp, meaning a second-order phase transition. Further, for $P>P_{c}$, the plot indicates monotonic behaviour. 

\begin{widetext}

\begin{figure}[h!h!] 
\begin{tabular}{c c}
\includegraphics[width=0.5\textwidth]{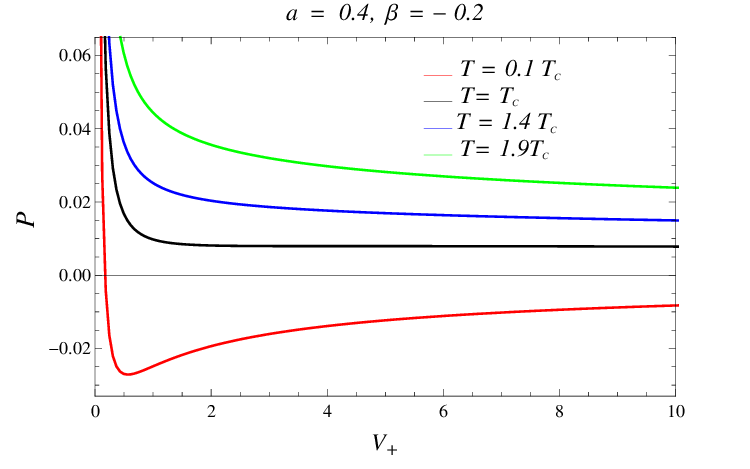}
\includegraphics[width=0.5\textwidth]{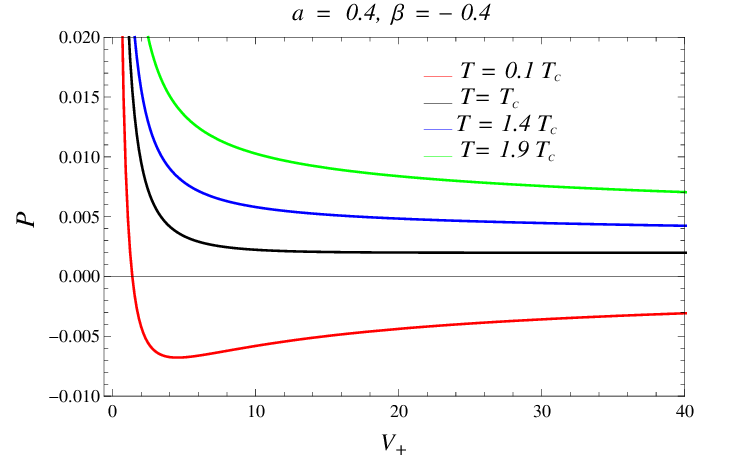}
\end{tabular}
\caption{The pressure $P$ vs volume $V_{+}$ plot for $a=0.4$, $\beta=-0.2$ (left) and $\beta=-0.4$ (right) showing the isotherms for Letelier spacetime immersed in perfect fluid dark matter. }
\label{fig : pvv}
\end{figure}
\end{widetext}


 In the limit $\gamma \to 0$ the specific heat and free energy reduce to those of nonsingular AdS black holes endowed with clouds of strings \cite{Sood:2022fio} and for $a=0$ we retrieve the heat capacity and Gibbs free energy for Schwarzschild AdS black holes \cite{Kubiznak:2012wp}.  
Next, we investigate the $P-V$ criticality of our solution via the equation of state 
\begin{equation}\label{state}
    P=\frac{T_{+}}{2r_{+}}-\frac{(1-a)}{r_+^2}-\frac{\beta}{8\pi r_{+}^3}.
\end{equation}
The various isotherms of Eq (\ref{state}) are shown in Fig \ref{fig : pvv}, which is indicative of a liquid-to-gas-like phase transition for values of temperature below the critical temperature ($T_{c}$). The plot shows ideal gas behaviour for isotherms with temperature values above the critical temperature $T_{c}$. The isotherms with high pressures correspond to LBH, and low-pressure isotherms show the SBH. The first-order phase transition corresponding to the Van der Waals theory is represented by the oscillating branch between SBH and LBH \cite{Kubiznak:2012wp}. The point of inflection is a solution of the equation.
\begin{equation}
\label{infl}
    \left.{\frac{\partial P}{\partial r_{+}}}\right\vert_{T_{+}}\,=\,0\,=\,\left.\frac{\partial^{2} P}{\partial r_{+}^{2}}\right\vert_{T_{+}}.
\end{equation}
The expression for critical temperature and pressure can be computed using Eq (\ref{infl}) 
\begin{equation}\label{Tcrit}
    T_{c} = \frac{3\beta+2r_{+}(1-a)}{4\pi r_{+}^2}
\end{equation}
\begin{equation}\label{Pcrit}
    P_{c}=\frac{3\beta\left[r_{+}(1-a)+\beta\right]\,+\,r_{+}^2\left(1-a\right)^2}{12\pi r_{+}^3 \beta}
\end{equation}

Inserting the expression for critical radius, we get the expressions for temperature, pressure, and volume 
\begin{equation}
r_{c}=-\frac{3\,\beta}{1-a}, \quad\quad T_{c}=-\frac{(1-a)^2}{12 \pi \beta},    
\label{rad}
\quad P_{c}=\frac{(1-a)^3}{216 \pi \beta^2},\quad\quad V_{c}=-\frac{36 \pi \beta^3}{(1-a)^3},   
\end{equation}
and the universal constant $ \epsilon= P_{c}V_{c}^{1/3}/T_{c}$ for Letelier AdS black holes in PFDM is $0.28$. In comparison to the value of $\epsilon$ for van der Waals gas, which is $0.375$ \cite{Kumar:2023gjt} the value of the universal constant is corrected for the solution under investigation. The critical values of the horizon radius $r_c$ depends on the cloud of strings parameter and the PFDM parameter and so are the other critical values.

\section{Phase transition behaviour of the black holes}
In this section, we study the phase structure of the Letelier AdS black hole endowed with PFDM during the first-order phase transition between the SBH and LBH. The whole study will be in the reduced parameter space. The coexistence curve in the $\tilde{P}-\tilde{T}$ plane may be parametrized in the dimensionless way
\begin{eqnarray}
\label{redcued_p}
\Tilde{P}&&=-28.2377 \tilde{T}^{10}+159.835 \tilde{T}^9-392.309 \tilde{T}^8+549.059 \tilde{T}^7-483.738 \tilde{T}^6+279.524 \tilde{T}^5-106.737 \tilde{T}^4\nonumber\\
&+&26.7808 \tilde{T}^3-3.51489 \tilde{T}^2+0.349711 \tilde{T}-0.0126105
\end{eqnarray}
where we define the dimensionless variables as 
\begin{equation}
    \tilde{T}=\frac{T}{T_c} \qquad \tilde{G}=\frac{G}{G_c} \qquad\tilde{P}=\frac{P}{P_c} \qquad \tilde{V}=\frac{V}{V_c}.
\end{equation}
The reduced forms of these variables, along with the other variables, determine the phase transition behaviour. As is evident from the first law itself  (\ref{firstlaw}), the mass of the black hole in AdS spacetime has a unique prescription, namely the enthalpy designated as $H$. This prescription leads us to define the Gibbs free energy, $\tilde{G}=\tilde{M}_+-\tilde{T}_+\tilde{S}_+$, which is defined in Eq.~\ref{gibbs}. 

\begin{widetext}

\begin{figure}[h!h!] 
\begin{tabular}{c c }
\includegraphics[width=0.5\textwidth]{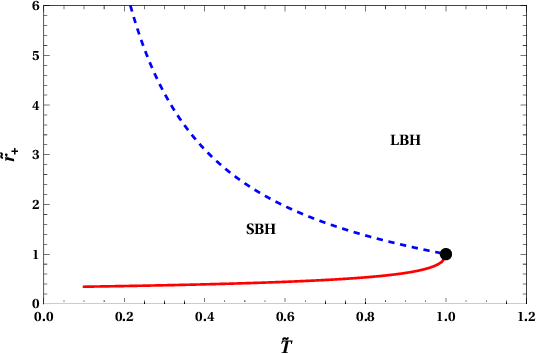}   \includegraphics[width=0.5\textwidth]{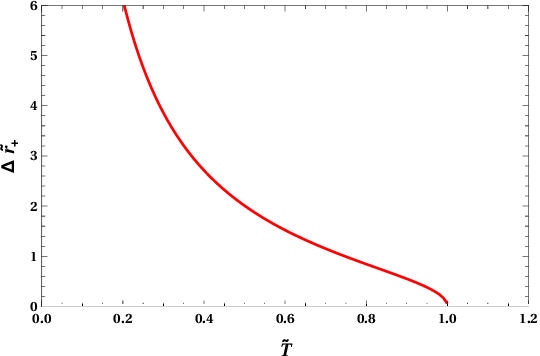}
\end{tabular}
\caption{Plot showing the variation reduced horizon radius $\tilde{r}_+$ vs the Hawking temperature $\tilde{T}$ (left) and differences of the reduced horizon radius $\Delta\tilde{r}$ vs the Hawking temperature (right). We see that the plot at the right has the value $\Delta\tilde{r}=0$ exactly at $\tilde{T=1}$ mimicking the critical point, and beyond it, there is a second-order phase transition.}
\label{fig :rtdrt}
\end{figure}
\end{widetext}
\begin{widetext}
\begin{figure}[h!] 
\begin{tabular}{c c }
    \includegraphics[width=0.5\textwidth]{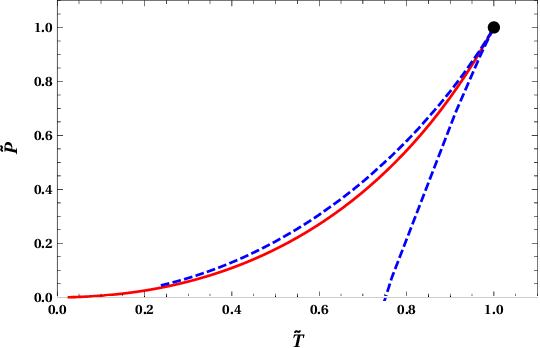}\hspace{-0.2mm}    \includegraphics[width=0.5\textwidth]{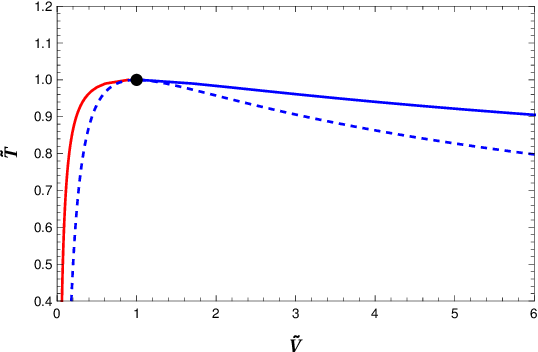}\\
\end{tabular}
\caption{Phase structure of Letelier AdS black holes in PFDM. In $\tilde{P}-\tilde{T}$ and $\tilde{T}-\tilde{V}$ diagrams the coexistence curve is shown in a solid line and the spinodal curves are shown in a dashed line.}
\label{fig:pttv}
\end{figure}
\end{widetext}

We have the critical horizon radius at the critical point, $r_{+c}$. Then the reduced horizon radius is expressed as $\tilde{r}_+=r_+/r_{+c}$. 
$\tilde{r}_+$ is a function of $(\tilde{P},\tilde{T})$ as is seen from the equation of state itself. We show the phase diagram in $(\tilde{P}-\tilde{T})$ plane, see Fig.~\ref{fig:pttv}. The black dot in the figure denotes the critical point \cite{Hegde:2021hjq}. We see from Fig.~\ref{fig:pttv} that in the $\tilde{P}-\tilde{T}$ plane, the metastable curve (the blue dotted line) for the small and the large black hole phases. Such a scenario dictates the fact that at the critical points, the phase transition terminates and just above it the phases are supercritical. In particular, for the coexistence curves, we have $\left(\Delta{G}\right)_{\Tilde{P},\Tilde{T}}=0$, whereas for the metastable curves we have the following conditions 
\begin{equation}
   ( \partial _V P)_T=0, \qquad (\partial _V T)_P=0.
\end{equation}

Therefore, the metastable curves are the locus of the extremal points. The region between the coexistence curve and metastable curve is the metastable phase as is depicted in the $(\tilde{T}-\tilde{S})$ or the $(\tilde{P}-\tilde{V})$ diagrams \cite{Hegde:2021hjq}. In the above analysis of the phase transition diagrams, we find that a connection should exist between the black hole horizon geometries and the phase transition behaviour of the black holes. Such distinction hints us to link the behaviour of the null geodesics or the photon orbits to the thermodynamic phase transition behaviour of the black holes. We can see from Fig~\ref{fig:pttv}, that the coexistence of small and large black holes shows the same behaviour in the $(\tilde{P}-\tilde{T})$ diagram, which for the metastable curves acts differently. Therefore, the thermodynamic notion for the metastable and the coexistence curves are not the same.

\section{Null geodesics of Letelier AdS black holes in PFDM.}
This section considers a free photon orbiting a black hole in the equatorial plane ($\theta=\pi/2$). The Lagrangian takes the form
\begin{equation}
2 \mathcal{L}=-f(r)\dot{t}^2+\frac{\dot{r}^2}{f(r)}+r^2\dot{\phi}^2,
\end{equation}
where the dots over the coordinates are meant for differentiation with respect to the proper time. For the Letelier AdS black holes endowed with PFDM there are two killing vectors $\partial_{t}$ and $\partial_{\phi}$ such that 
\begin{eqnarray}
p_t=-f(r)\dot{t}\equiv E\\
p_\phi= r^2\dot{\phi} \equiv L\\
p_r=\dot{r}/f(r).
\end{eqnarray}
which correspond to the conserved quantities $E$ and $L$ along the geodesics. The quantities $E$ and $L$ are the photon's energy and orbital angular momentum, respectively, denoting the constants of motion. The equations of motion for the $t$ and $\phi$ coordinates read
\begin{equation}
\dot{t}=\frac{E}{f(r)}
\end{equation}
\begin{equation}
\dot{\phi}=\frac{L}{r^2 \sin ^2\theta}.
\end{equation}
The Hamiltonian for the black holes
\begin{equation}
2\mathcal{H}=-E\dot{t}+L\dot{\phi}+\dot{r}^2/f(r)=0.
\end{equation}
We are interested only in the radial motion. The radial equation of motion is obtained as, 
\begin{equation}
\dot{r}^2+V_{eff}=0
\end{equation}
where $V_{eff}$ is the effective potential. We define the dimensionless effective potential $\tilde{V}_{eff}=V_{eff}/E^2$, which has the following explicit form
\begin{equation}
\tilde{V}_{eff}=\frac{u^2}{r^2}f(r)-1.
\end{equation}
The quantity $u=L/E$ in the above equation is called the impact parameter.
\begin{figure}[h!h!] 
\begin{tabular}{c c }
    \includegraphics[width=0.5\textwidth]{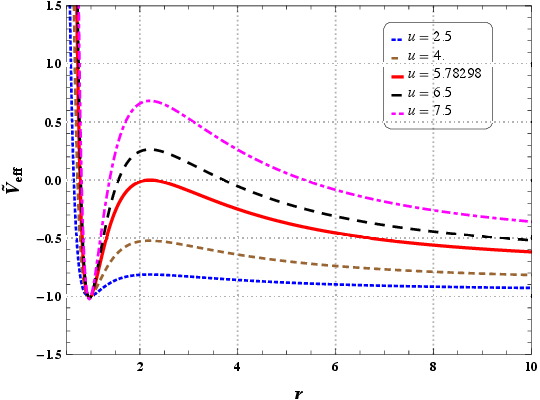}
\end{tabular}
\caption{The effective potential for Letelier AdS black holes in PFDM.}
\label{fig :effective}
\end{figure}
We have plotted the behavior of the effective potential $\tilde{V}_{eff}$ with the radial coordinate as can be seen from Fig. \ref{fig :effective} for different values of impact parameter $u$, for three different cases--\\
\[\textbf{a.}\;u>u_c,\; \text{the unbounded motion}\] \[\textbf{b.}\;u=u_c,\;\text{the critically bounded}\] and\[\textbf{c.}\;u<u_c,\;\text{for the impinging one}\]
The only accessible region for the photon is $V_{eff}<0$, since $\dot{r}^2>0$ there. As is evident from Fig.~\ref{fig :effective}, the photon falls into the black hole for $u<u_c$, while for $u>u_c$ it is reflected off the black holes. Between these two cases, there is a critical case $u=u_{c}=5.78298$ which is shown by a red thick line (see Fig.~\ref{fig :effective}). The condition \textbf{c} is meant for the unstable photon orbits, and they are circular. The angular momentum of the photon becomes critical at the point where $\tilde{V}_{eff}=0$ (red solid curve) as is shown in Fig.~\ref{fig :effective}. As a consequence of the effective potential to be zero, the radial velocity of the photon also vanishes there. The value of the radial coordinate at the point where the radial velocity is zero is called the photon orbit radius. The unstable circular orbits are determined through the relations,
\begin{equation}
V_{eff}=0\quad , \quad V'_{eff}=0 \quad , \quad V''_{eff}<0,
\label{unstable}
\end{equation}
where prime denotes the differentiation with respect to the radial coordinate $r$. From the second equation $V'_{eff}=0$, we find the photon orbit radius $r_{ps}$ satisfies
\begin{equation}
2f(r_{ps})-r_{ps}\partial _r f(r_{ps})=0,
\label{aneqn}
\end{equation}
which gives the radius of photon sphere $r_{ps}$ as

\begin{equation}
r_{ps}=-\frac{3 \beta W\left(\frac{2}{3} (a-1) e^{\frac{2 M}{q}+\frac{1}{3}}\right)}{2 (a-1)},
\label{rpsDM}
\end{equation}
whose peak approaches zero where the radial velocity of the photon vanishes and ``W'' is the Lambert $W$ function. Given the parametric values $a>0$ and $\beta<0$, we must restrict ourselves to the output of the Lambert $W$ function to be negative as $r_{ps}>0$ always.

The above photon orbit radius is used to calculate the minimum impact parameter. Solving $(\tilde{V}_{eff}=0)$ to obtain,
\begin{equation}
u_{ps}=\frac{L_c}{E}=\left. \frac{r}{\sqrt{f(r)}} \right| _{r_{ps}}.
\label{upsequation}
\end{equation}
The explicit form of this can be obtained 
\begin{equation*}
  u_{ps}=\frac{\left(3\beta\mathcal{Y}\right)^{3/2}}{(1-a)^3\left[-4M+3\beta\mathcal{Y}+2\beta\log\left(\frac{3\mathcal{Y}}{2}\frac{|\beta|}{(1-a)}\right)+18P\pi \beta^3\mathcal{Y}^3\right]}  
\end{equation*}

with
\begin{equation}
    \mathcal{Y}=W\left(\frac{2}{3} (a-1) e^{\frac{2 M}{\beta}+\frac{1}{3}}\right).
\end{equation}
For a photon coming from a distant source, it encounters strong gravity effects while passing by the black holes. The photon would experience a maximum deflection as $u\to u_{ps}$, where the deflection angle is infinitely large. Such a phenomenon is described by black hole lensing observables. Therefore, for the photon with large $u$, the deflection angle is smaller, but it will get more deflected as $u$ decreases, and the deflection becomes maximum at $u=u_{ps}$, where the photons are unbounded. Therefore, the quantities $r_{ps}$ and $u_{ps}$ play essential roles in describing the photon spheres. In the subsequent discussion, we investigate the behaviour using the photon sphere parameters $r_{ps}$ and $u_{ps}$ and find that there should exist a correlation to the small-large black hole phase transition. The critical quantity which plays a vital role in such a connection is the mass parameter $M$. We determine the mass in terms of lensing variables $r_{ps}$ and $u_{ps}$ and use them to express in terms of the reduced pressure $\tilde{P}$ and the reduced horizon radius $\tilde{S}$. This way we can find the correlation of the lensing observables to the thermodynamic variables as seen from Fig.~\ref{fig :r_t-u_t}. 
\subsection{Critical behaviour of the photosphere radius and the minimum impact parameter}
Our main concern is to study the behaviour of the $r_{ps}$ and $u_{ps}$ along the isobars and isotherms. A straightforward analysis leads us to numerically study the reduced temperature $\tilde{T}$, as a function of the reduced photon orbit radius $r_{ps}$, and the reduced minimum factor $u_{ps}$. Figs.~\ref{fig :r_t-u_t} and \ref{fig :T_rps-T_ups} show that for smaller values of the temperatures, we have smaller values of both the quantities $r_{ps}$ and $u_{ps}$, which at larger temperatures become larger. However, the behavior for the reduced pressure $\tilde{P}<1$, has interesting features. In such cases, we have similar behaviours of the isobaric curves as in the usual vdW-like fields in the $\tilde{T}-\tilde{S}$ plane in the ordinary thermodynamic systems. If one goes on increasing the $\tilde{P}$ values and reaches $\tilde{P}=1$, the extrema of the curves in $\tilde{T}-\tilde{r}_{ps}$ plane coincide. As an obvious conclusion, for $\tilde{P}>1$, we have the isobars which increase monotonically. Similar phenomena are observed in the numerical analysis of the $\tilde{T}-\tilde{u}_{ps}$ diagrams. We have also plotted the diagrams in the  $\tilde{P}-\tilde{r}_{ps}$ and $\tilde{P}-\tilde{u}_{ps}$ planes. All such plots for $\tilde{T}<1$ correspond to the thermal phase transition of the first order, while at $\tilde{T}=1$, we have the critical points, thereby representing a second-order phase transition. We have no phase transitions for $\tilde{T}>1$. However, whether such phase transitions would emerge from Maxwell's equal area law is not clear in the case of photon orbits or minimum impact parameters. One should not forget to mention that such curves depict the isothermal curves in the $\tilde{P}-\tilde{V}$ plane of the ordinary thermodynamic system, a case study of course. Having experienced such features of the usual vdW-like fields, the mimickers of the isothermal or isobaric curves in the $\tilde{T}-\tilde{r}_{ps}(\tilde{u}_{ps})$ or $\tilde{P}-\tilde{r}_{ps}(\tilde{u}_{ps})$ planes must have some connections to the thermodynamic phase transitions via the small-large black hole phases.

\begin{widetext}
\begin{figure}[h!h!] 
\begin{tabular}{c c }
    \includegraphics[width=0.45\textwidth]{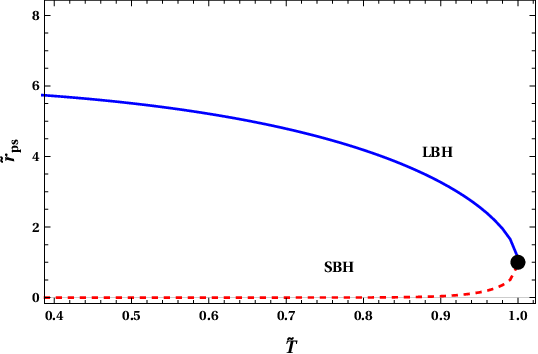}
    \includegraphics[width=0.45\textwidth]{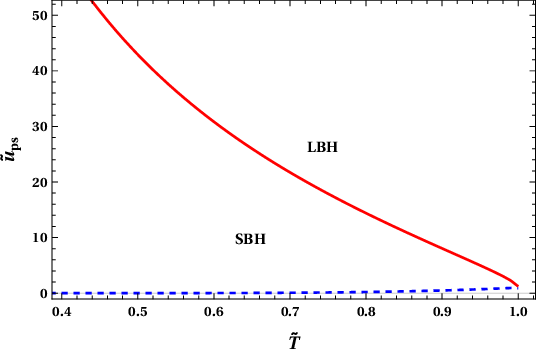}
\end{tabular}
\caption{Reduced photon orbit radius vs the Hawking temperature}
\label{fig :r_t-u_t}
\end{figure}

\begin{figure}[h!h!] 
\begin{tabular}{c c }   \includegraphics[width=0.45\textwidth]{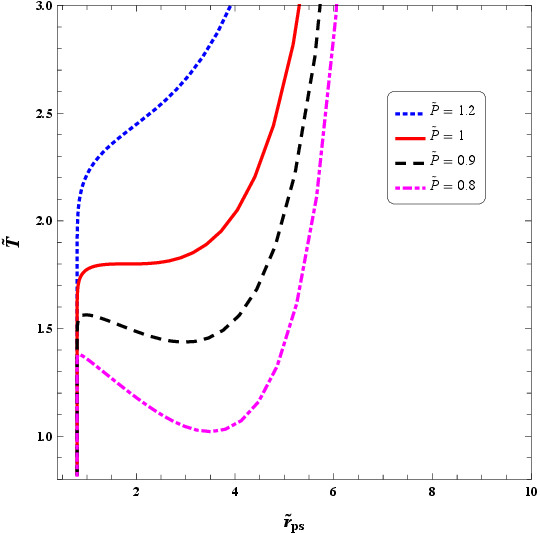}    \includegraphics[width=0.45\textwidth]{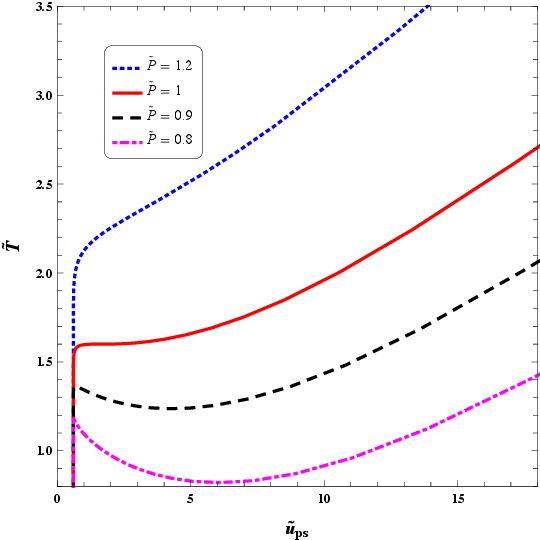}
\end{tabular}
\caption {Reduced Hawking temperature ($\tilde{T}$) : (a) $\tilde{T}$ vs $\tilde{r}_{ps}$ (left)  (b) $\tilde{T}$  vs $\tilde{u}_{ps}$ (right).}
\label{fig :T_rps-T_ups}
\end{figure}

\begin{figure}[h!h!] 
\begin{tabular}{c c }   \includegraphics[width=0.45\textwidth]{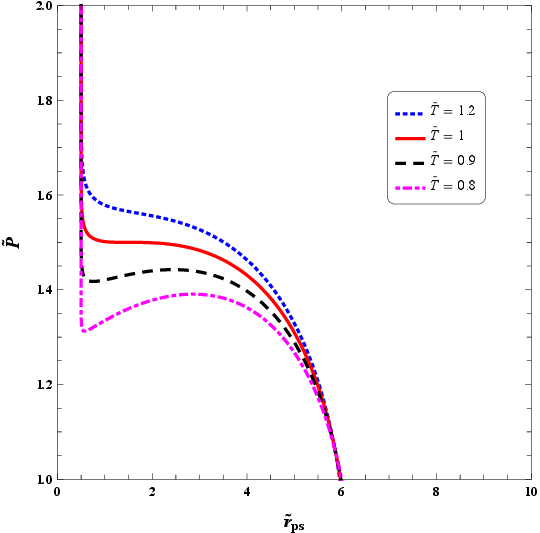}   \includegraphics[width=0.45\textwidth]{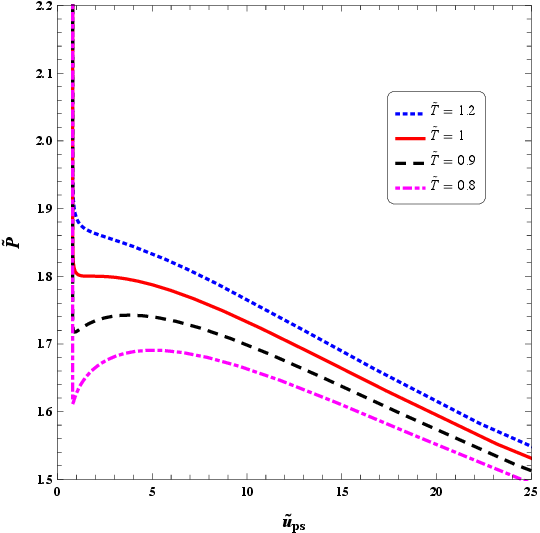}
\end{tabular}
\caption{Reduced pressure ($\tilde{P}$) : (a) $\tilde{P}$ vs $\tilde{r}_{ps}$ (left) (b) $\tilde{P}$ vs $\tilde{u}_{ps}$ (right).}
\label{fig :P_rps-P_ups}
\end{figure}
\begin{figure}[h!h!] 
\begin{tabular}{c c }
    \includegraphics[width=0.5\textwidth]{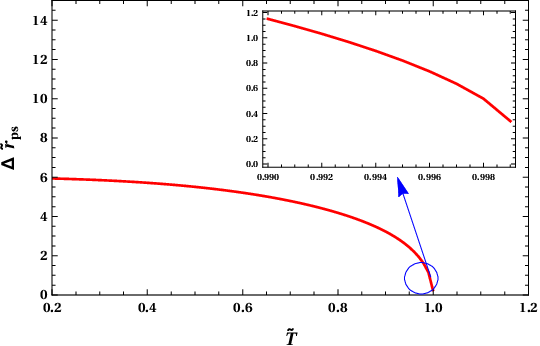}
    \includegraphics[width=0.5\textwidth]{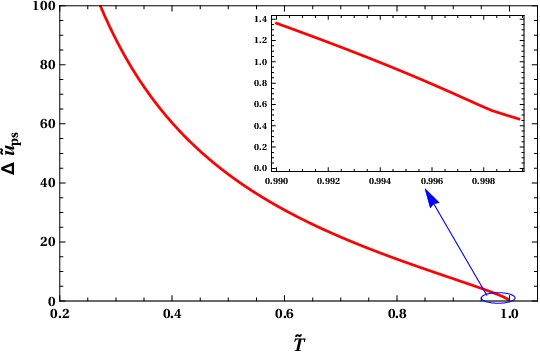}
\end{tabular}
\caption{Variation in reduced photon orbit radius $\tilde{r}_{ps}$ (left) and minimum impact factor $\tilde{u}_{ps}$ (right) vs the Hawking temperature $\tilde{T}$.}
\label{fig :delta_r_t-u_t}
\end{figure}
\end{widetext}
 We show the plot for $\tilde{r}_{ps}$ or $\tilde{u}_{ps}$, as a function of phase transition temperature $\tilde{T}$ (see the Fig.~\ref{fig :T_rps-T_ups}, for your reference). As we increase the temperature $\tilde{T}$, the value in the photon orbit radius $\tilde{r}_{ps}$ decreases for the coexistence large black hole. On the contrary, for the increase in the values of the quantity $\tilde{T}$, the radius $\tilde{r}_{ps}$ increases for the coexistence of small black holes. For the phase transition temperature, $\tilde{T}=1$, the radius $\tilde{r}_{ps}$ approaches the same value for the coexistence of small and large black holes. 
 \\
 \begin{widetext}
\begin{figure}[h!h!] 
\begin{tabular}{c c }
    \includegraphics[width=0.5\textwidth]{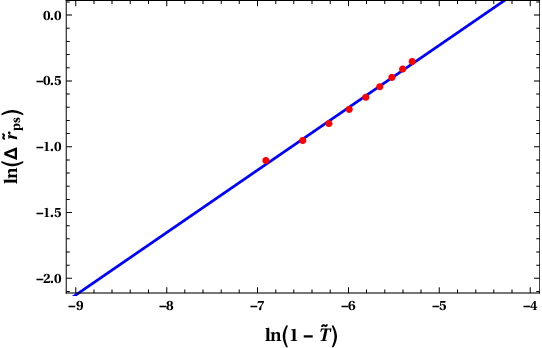}
    \includegraphics[width=0.5\textwidth]{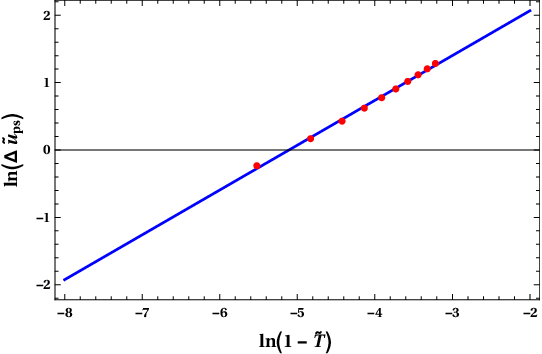}
\end{tabular}
\caption{Fitting curves of reduced photon orbit radius $\tilde{r}_{ps}$ (left) and minimum impact factor $\tilde{u}_{ps}$ (right) vs the Hawking temperature $\tilde{T}$. }
\label{fig :delta_r_t-u_t}
\end{figure}
\end{widetext}
 We also plot the relative differences $\Delta\tilde{r}_{ps}/\tilde{r}_{ps}$ or $\Delta\tilde{u}_{ps}/\tilde{u}_{ps}$ with respect to the phase transition temperature. As expected, the differences decrease with increasing values of the phase transition temperature, and ultimately it vanishes at the critical point. The interesting thing that may be a consequence of such results is that the quantity $\Delta\tilde{r}_{ps}/\tilde{r}_{ps}$ or $\Delta\tilde{u}_{ps}/\tilde{u}_{ps}$ behaves as the order parameters. They correspond to the non-vanishing contribution for the first-order phase transition and become zero at the critical point, and the phase transition is second order. The critical exponent around the critical points will simply follow the relation 

\begin{eqnarray}
\label{criticalexp}
\Delta\tilde{r}_{ps}/\tilde{r}_{ps},\;\; \Delta\tilde{u}_{ps}/\tilde{u}_{ps}\sim a \left({1-\tilde{T}}\right)^{\delta}
\end{eqnarray}
which is also written as
\begin{eqnarray}
\label{criticalexp}
\ln\left({\Delta\tilde{r}_{ps}/\tilde{r}_{ps}}\right),\;\; \ln\left({\Delta\tilde{u}_{ps}/\tilde{u}_{ps}}\right)\sim \ln a+\delta \ln\left({1-\tilde{T}}\right)
\end{eqnarray}
The logarithmic variation of $\Delta{\tilde{r}_{ps}}/\tilde{r}_{ps}$ or $\Delta{\tilde{r}_{ps}}/\tilde{r}_{ps}$ is linear with respect to $1-\tilde{T}$ which helps us in determining the critical exponents. A numerical fit to the logarithmic variation yields the output $\delta$ for $\Delta{\tilde{r}_{ps}}/\tilde{r}_{ps}$ to be $\delta=0.500918203$ or for $\Delta{\tilde{r}_{ps}}/\tilde{r}_{ps}$ $\delta=0.490567809$. Like a wide class of black holes in asymptotically AdS spacetime in four as well as in the higher dimensions, for our black hole too the approximate value of the critical exponent turns out to be $\delta\sim1/2$ at the critical points. 

\section{Conclusion}
We have examined the thermodynamics $P-V$ criticality and the relationship between photon orbit radius and the thermodynamic phase transition for Letelier AdS black holes in PFDM which may be useful to link gravity and thermodynamics of black holes. First, we obtained an exact Letelier AdS black hole solution in PFDM with additional clouds of strings parameter ($a$) and PFDM parameter ($\beta$) and discussed the energy conditions. Next, we extend the extended phase space thermodynamics of Letelier AdS black holes in PFDM, showing an interesting phase transition. We have shown that the thermodynamic $ansatz$ viz, ADM mass, Hawking temperature, specific heat and free energy are corrected overall by the additional parameters $a$ and $\beta$. The behaviour of Letelier AdS black holes with PFDM is consistent with the van der Waals fluid. $\tilde{G}$ vs $\tilde{T}$ analysis confers first-order phase transition. The analysis of heat capacity dictates a second-order phase transition. We also used the mass term as a function of the thermodynamic variables in the reduced parameter space. 

We examined the relationship between the photon sphere radius and the thermodynamic phase transition for the  Letelier AdS black hole immersed in PFDM. This will provide a new way to realize the link between the gravity and thermodynamics of black holes.  We found that the isobaric and the isothermal curves concerning the photon orbit radius $\tilde{r}_{ps}$, and also with the minimum impact parameter $\tilde{u}_{ps}$, in the reduced parameter space, signify the same thermal phase transition behaviour of the usual van der Waals like fluids. This means that the curves in the  $\tilde{T}-\tilde{r}_{ps}(\tilde{u}_{ps})$ or in the $\tilde{P}-\tilde{r}_{ps}(\tilde{u}_{ps})$ planes show the oscillatory behaviours confirming the first order phase transition of the van der Waals fluids. The oscillatory behaviour confirms that two extrema should exist. When the two extrema coincide, we have the second-order phase transition. For other parameters above the second-order phase transition values, we do not have any such curves with a higher-order phase. We also plotted the relative differences $\Delta{r_{ps}}/r_{ps}$ and $\Delta{u_{ps}}/u_{ps}$, to show that they may serve as the order parameters near the critical points. We have the critical exponent $\delta\sim 1/2$ at the critical points. This value of the critical exponent is the universal feature of many black hole systems in four and the higher dimensional spacetimes in asymptotically AdS spaces. 

A groundbreaking development has emerged wherein the AdS black hole is depicted dually as a thermal system, specifically through the holographic counterpart of the extended phase space thermodynamics. This breakthrough stems from the perturbation of the conformal factor within the dual CFT metric, as outlined in \cite{Ahmed:2023snm}. Our aspiration is for our research to be expanded upon within the realm of holographic thermodynamics. It holds substantial promise to explore analogous analyses within different spacetime geometries, including but not limited to the de Sitter black hole, the black hole solution within alternative gravitational frameworks, or those arising from higher curvature theories.
\section{Acknowledgement}
 This work was supported by the National Natural Science Foundation of China
(Grants No. 11875151, No. 12347177 and No. 12247101), the 111 Project under (Grant No. B20063) and Lanzhou City's scientific research funding subsidy to Lanzhou University.

\end{document}